# Effect of Macroscopic Surface Heterogeneities on an Advancing Contact Line.

S. S. Melides[1,2], D. Vella[3]. M. Ramaioli[2,\*]

[1] Department of Chemical and Process Engineering, University of Surrey, Guildford, United Kingdom
[2] UMR SayFood, Université Paris-Saclay, AgroParisTech, INRAE, 78850, Thiverval Grignon, France
[3] Mathematical Institute, University of Oxford, Woodstock Rd, Oxford, OX2 6GG, UK
[\*] *marco.ramaioli@inrae.fr*

## Abstract

The shape of a liquid-air interface advancing on a heterogeneous surface was studied experimentally, together with the force induced by the pinning of the contact line to surface defects. Different surfaces were considered with circular defects introduced as arrays of cocoa butter patches or small circular holes. These heterogeneous surfaces were submerged into aqueous ethanol solutions, while measuring the additional force arising from the deformation of the advancing contact line and characterising the interface shape and its pinning on the defects. Initially, the submersion force is linear with submerged depth, suggesting a constant defect-induced stiffness. This regime ends when the contact line de pins from the defects. A simple scaling is proposed to describe the depinning force and the depinning energy. We find that, as the defect separation increases, the interface stiffness increases too, with a weak dependency on the defect radius. This interaction between defects cannot be captured by the simple scaling, but can be well predicted by a theory considering the interface deformation in presence of periodic arrays of holes.
Creating a 4-phase contact line, by including solid defects (cocoa butter) reduced pinning forces. The radius of the defect had a nonlinear effect on the depinning depth. The 4-phase contact line resulted in depinning before the defects are fully submerged.
These experimental results and the associated theory help understanding quantitatively the extent to which surface heterogeneities can slow down wetting. This in turn paves the way to tailor the design of heterogeneous surfaces, towards desired wetting performances.

## Introduction

It is well known that wetting of heterogeneous food powders is hindered by the presence of fat on the powder surface. Typically, manufacturers aim to reduce surface fat to improve wettability. Previous research[1-5] has focused on the wettability of homogenous insoluble films of varying hydrophilicity neglecting the chemical surface heterogeneities that are present on most real food powders. Other studies[6-11] considered the bulk wetting and dispersion properties of real powders, but have not characterised and elucidated the role of microscopic heterogeneities on the powder surface.

In the first instance, we are interested in model systems in which one can evaluate the effect of defect topology and periodicity on the moving contact line deformation and on the force needed to overcome the defects. The spontaneous spreading of liquids on surfaces occurs to reduce the total surface energy on the different interfaces, as described by the Young equation. The relative magnitude of the specific energy of a wetted and unwetted solid surface, can be evaluated by considering the spreading factor (S)[12]:

$$S = \gamma_{SG} - (\gamma_{SL} + \gamma) = \gamma(cos\theta - 1)$$

Equation 1

Where $\gamma$ indicates the liquid surface tension (liquid-air interface), $\gamma_{SG}$ and $\gamma_{SL}$ are the specific surface energies of the solid-gas interface and of the solid-liquid interface respectively. For partially wetting surfaces, $\theta$ represents the equilibrium contact angle.

For most real surfaces, the contact line can be affected by surface defects, leading to pinning and a deviation from the Young's equilibrium contact angle[12-22]. The microscopic contact angle is dependent on the local chemical surface composition and can vary along the contact line. As the contact line moves, it will deform over the different defects to minimise the total energy at the interface. At the macroscopic scale, the apparent contact angle will adjust to account for the microscopic pinning caused by defects. The contact angle will be decreased by the presence of high energy (hydrophilic) defects on low energy (hydrophobic) surface whilst a contact line recedes; the reverse (an increase of the apparent contact angle for low energy defects on high energy surfaces) is true for advancing contact lines[19, 23-25]. The advancing contact angle ($\theta_{adv.}$) then is increased from $\theta$ for real advancing contact lines. Likewise, the receding contact lines can give rise to a reduced apparent contact angle termed the receding contact angle ($\theta_{rec.}$). The difference between $\theta_{adv.}$ and $\theta_{rec.}$ is known as the contact angle hysteresis, and its magnitude depends on multiple factors, including defect size, shape, frequency/distribution and chemical composition[12, 26]. As the defects become bigger and more dispersed, the macroscopic contact line can exhibit an intermittent motion, whereby the contact line will first pin, and then de-pin from the defect. When a moving contact line encounters a defect, a free energy barrier needs to be overcome before the contact line will move across the defect[15, 27].

There exist two types of defect: chemical heterogeneities and physical defects (topological). Equation 1 is affected by the presence of a chemical defect. Previous authors[26] have proposed to consider the difference ΔS (Equation 2) as the change in energy per unit interface caused by chemical defects. The resulting force generated by a defect is expressed in Equation 3.

$$\Delta S = \Delta\gamma\left(cos\theta_{def} - cos\theta_{sub}\right)$$

Equation 2

$$f = 2r\Delta S$$

Equation 3

Similarly, the energy to overcome a defect can be calculated by multiplying the change in spreading factor by the defect area ($E = \pi r^2 \Delta S$ for circular defects).

The maximum equilibrium contact angle from chemical heterogeneities is thought to be less than 120° [28, 29]. However, much higher contact angles have been observed over real surfaces, because of geometrical (topological) surface heterogeneities[30-34]: defects allow air to be trapped on the surface as first described by Cassie and Baxter[35].

Defect density also affects the contact line shape. For small defects (e.g. <10 µm), dense packing has been shown to lead to a macroscopic contact line that is unperturbed[15] and indistinguishable for a straight line. As the defect size increases the perturbations in the contact line shape become observable. It has been proposed that neighbouring defects synergistically affect the contact line when their distance is lower than the capillary length ($\kappa^{-1} = (\gamma/\rho g)^{1/2}$)[12, 26]. Beyond this critical distance gravity is assumed to screen defects from one another.

As yet, the effect of an array of defects on the deformation of a contact line (and interface), as well as the effect of this deformation on the resulting force is not clearly understood. This study seeks to bridge this gap. Similarly to other authors[7, 15, 16, 19-21] we investigate the effects of defects on moving contact lines on vertical substrates, to include the effect of gravity on the capillary / meniscus rise. This allows for the use of large macroscopic defects (>500µm), which makes it easier to evaluate the synergistic effects of defects on the contact line, both in terms of the interfacial deformation and the resulting forces. We propose a theory predicting the contact line shape and pinning force for small inclines of the surface. We compare the model with experimental data both within and beyond the small angle range, characterising also the depinning condition.

## Theory

This paragraph presents the theoretical description of a planar liquid interface that is deformed by the entry of a plate decorated by circular defects. The plate enters the liquid normal to the interface and parallel to the direction of gravity. The theory considers only one of the two symmetric interfaces created by the immersion of the plate.

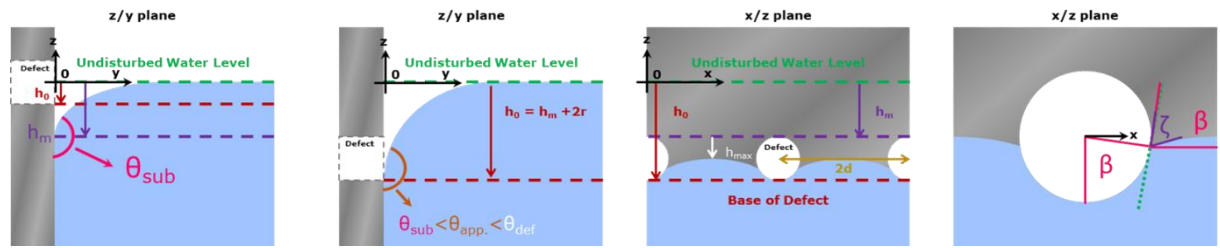

Figure 1. Schematics of the key variables used to describe the geometry and position of the defects submerged into a liquid.

The geometry of the system studied is illustrated in Figure 1. An array of defects of radius $r$ and centre-centre spacing $2d$ is submerged in a liquid. The position of the defects is characterised by the distance of the bottom of the defects from the undisturbed liquid interface ($h_0$). Cartesian coordinates ($x,y,z$) centred with the centre of a hole are introduced with the ($x,y$) plane parallel to the undisturbed interface and the $z$ axis normal to the interface. In the following, lower case symbols are used to denote dimensional variables, while upper case symbols are used to denote dimensionless variables.

The deflection of the liquid-air interface from its undisturbed level, denoted $h(x,y)$, satisfies the Laplace-Young Equation[17]:

$$\rho g h = \gamma \frac{\nabla^2 h}{[1 + |\nabla h|^2]^{3/2}}$$

Equation 4

Focusing on the case of linear deformations $|\nabla h|^2 \ll 1$, and normalizing using the capillary length, $\kappa^{-1}$, Equation 4 becomes:

$$H = \nabla^2 H$$

Equation 5

We consider a periodic array of holes of dimensionless radius $R = r\kappa$, with horizontal period $2D = 2d\kappa$, that deforms the equilibrium contact line (CL) as indicated schematically in Figure 1. Because of the periodicity of the problem, we seek to solve Equation 5 on the domain $0 < X < D$, $0 < Y < \infty$, with symmetry imposed at $X = 0$. Boundary conditions on Equation 5 must therefore be imposed at $Y = 0$, and as $Y \to \infty$ (the direction coming out of the page in figure 1). The nature of the boundary conditions along the plate, $Y = 0$, depends on whether the contact line is pinned to the boundary of the hole (which occurs for $0 \leq X \leq R \sin \beta$, *defining the angle of depinning* $\beta$) or is in contact with a smooth part of the plate ($R \sin \beta \leq X \leq D$) in which case it is expected to satisfy a contact angle condition. If the vertical position of the base of the hole (relative to the undisturbed liquid level is denoted $H_0$ and the equilibrium contact angle by $\theta$ then we may take the boundary condition along the wall as:

$$H(X, 0) = H_0 + R\left(1 - \sqrt{1 - X^2/R^2}\right) \quad \text{for} \quad 0 \leq X \leq R \sin \beta$$

Equation 6

and

$$\left.\frac{\partial H}{\partial Y}\right|_{(X,0)} = -\cot\theta \quad \text{for} \quad R \sin \beta \leq X \leq D$$

Equation 7

The remaining boundary conditions are obtained from the symmetry inherent in the periodic boundary conditions at $X = 0$ and $X = D$ i.e.:

$$\left.\frac{\partial H}{\partial Y}\right|_{(0,Y)} = \left.\frac{\partial H}{\partial Y}\right|_{(D,Y)} = 0$$

Equation 8

and the condition that the interface returns to its natural level far from the plate, i.e.:

$$H \to 0 \quad as \quad Y \to \infty$$

Equation 9

Once the meniscus profile $H(X,Y)$ has been determined, the vertical force (f) acting on the plate from the meniscus can be computed as:

$$F = \frac{f}{\gamma \kappa^{-1}} \approx \int_0^D \left.\frac{\partial H}{\partial Y}\right|_{(X,0)} dX$$

Equation 10

We use a spectral collocation method[36] (described in the Appendix) to solve this problem numerically.

# Materials

Table 1 describes the experiments performed in this study. Aqueous solutions of EtOH were prepared at 0%, 5%, 30% and 50% w/w and used to assess the wetting with solid heterogeneous surfaces with different defects. In a first set of experiments, holes were drilled into brushed stainless steel (s/s) plates (dimensions of 21.8x20.0x0.9mm) to create defects that would induce interface pinning during wetting and a depression of the 3-phase contact line (CL). Hole radii ($r$) of 0.27mm and 0.56mm (±0.01 mm) were considered, in evenly spaced arrays of 4, 6, 8 and 12 defects leading to a centre-centre periodicity ($2d$) of 5.4, 3.6, 2.7 and 1.8 mm. Another set of plates was used, in which the defects were not aligned but staggered by a vertical offset ($o$) of 0.2 mm as described in Figure 4 D.

*Table 1. Experimental conditions considered in this study.*
*The first column indicate the label attributed to the liquid-solid pair considered. The first letter of the label indicates the nature of the substrate (S or G for glass or steel), the second whether the defect is a hole or made of cocoa butter (h or c), the third letter whether the defect is small, medium or large (S, M or L), the fourth letter whether the defects are inline or offset (i or o). The label ends with the digits 0, 30 or 50 depending on the EtOH w/w concentration used. For instance, label **ShSi 30** indicates a **S**teel substrate with **h**oles that are **S**mall and **i**nline wetted by a **30**% EtOH w/w solution. For convenience, italic is used in the table for the characters used to form the labels.*
*For all properties, the standard deviation and number of repetitions has been presented has been indicated in brackets. The 2$^{nd}$ and 3$^{rd}$ columns indicate the substrate used (S stands for stainless steel, G for glass) and the advancing contact angle ($\theta_{sub}$) of the liquid indicated in column 8 with the substrate (±8°, 8 repetitions). Columns 4-7 indicate respectively the type of defects (h-holes, c-cocoa butter), the defect radius (± 0.01mm, 30 repetitions) and whether this size was labelled small, medium or large, the defect-liquid advancing contact angle ($\theta_{def}$, ±8°, 8 repetitions, it is assumed 180° for holes) and the defect alignment (i- inline, o- offset, i\* indicates manually pipetted in a straight line). Columns 8-9 indicate the aqueous EtOH concentrations (w/w) and the corresponding surface tension ($\gamma$) (±0.06N/m, 3 repetitions). The last column indicates the difference of the substrate and defect spreading factor ΔS calculated as described in Eq.2 from the values in columns 3,6 and 9.*

| Sample | Substrate | | Defect | | | | Liquid | | ΔS=γΔcos$\theta_{adv.}$, N/m |
|---|---|---|---|---|---|---|---|---|---|
| | Mat. | $\theta_{sub.}$, ° | Type | r, mm | $\theta_{def.}$, ° | Alignment | EtOH, %w/w | $\gamma$, N/m | |
| *ShSi 0* | S | 95 | h | 0.27 *S* | 180 | *i* | 0 | 0.073 | -0.067 |
| *ShSo 0* | S | 95 | h | 0.27 *S* | 180 | *o* | 0 | 0.073 | -0.067 |
| *ShSi 30* | S | 70 | h | 0.27 *S* | 180 | *i* | 30 | 0.034 | -0.046 |
| *ShSi 50* | S | 50 | h | 0.27 *S* | 180 | *i* | 50 | 0.029 | -0.048 |
| *ShLi 0* | S | 95 | h | 0.56 *L* | 180 | *i* | 0 | 0.073 | -0.067 |
| *ShLo 0* | S | 95 | h | 0.56 *L* | 180 | *o* | 0 | 0.073 | -0.067 |
| *ShLi 30* | S | 70 | h | 0.56 *L* | 180 | *i* | 30 | 0.034 | -0.046 |
| *ShLi 50* | S | 50 | h | 0.56 *L* | 180 | *o* | 50 | 0.029 | -0.048 |
| *ScLi 0* | S | 95 | c | 0.56 *L* | 117 | *i* | 0 | 0.073 | -0.027 |
| *ScLi 30* | S | 70 | c | 0.56 *L* | 118 | *i* | 30 | 0.034 | -0.028 |
| *ScLi 50* | S | 50 | c | 0.56 *L* | 86 | *i* | 50 | 0.029 | -0.017 |
| *GcS 30* | G | 30 | c | 0.62 *S* | 118 | *i\** | 30 | 0.034 | -0.046 |
| *GcS 50* | G | 25 | c | 0.62 *S* | 86 | *i\** | 50 | 0.029 | -0.024 |
| *GcM 0* | G | 35 | c | 0.82 *M* | 117 | *i\** | 0 | 0.073 | -0.093 |
| *GcL 0* | G | 35 | c | 1.44 *L* | 117 | *i\** | 0 | 0.073 | -0.093 |

In a second set of experiments, cocoa butter (cb) (Barry Callebaut, Switzerland), was also used to fill the holes, creating hydrophobic defects. Because of the two different solid surfaces, these experiments created a deformed 4-phase CL.

In a third set of experiments, cocoa butter was deposited with a pipette on glass slides to create patches having the form of spherical cap, with radii 0.62 (±1.0),

0.81 (±0.08) and 1.41 (±0.16) mm and base angle of approximately 30°. This shape resulted in an increase in surface area of approximately 6.5%. Fat droplets were either deposited in rows of 1, 3, 4, 5, 6, or 8 for the mediums sized patches or 1, 3, 5 for the small and large.

## Methods

The radii and periodicity of the defects were verified by using an AXI reflectance microscope (Zeiss, Germany). The images were analysed using ImageJ.

Contact angles were measured by imaging droplets deposited on the selected surface using a high-definition camera (Basler, Germany). Images were then analysed using the 'Drop Snake' plug-in[37] for ImageJ to obtain the contact angle.

The surface tension was measured with a K12 tensiometer (Krüss, Germany) using a platinum Wilhelmy plate.

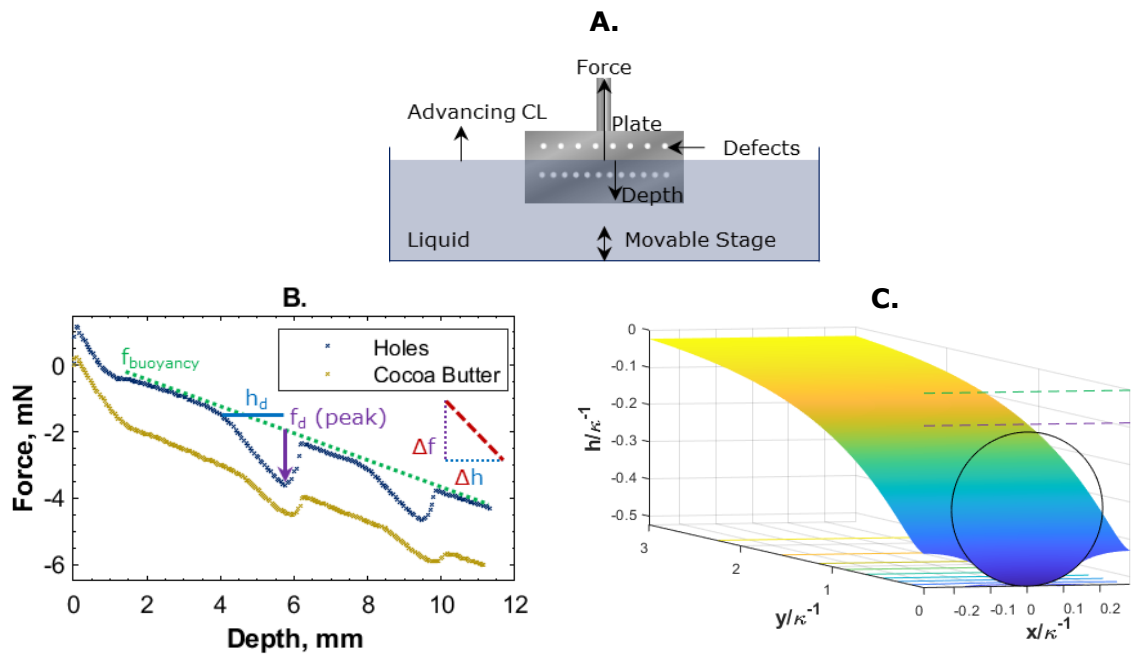

Figure 2. A. Schematic of experiment B. Typical force vs depth data as obtained for 12 then 8 defects. Results for r=0.56 mm holes and cocoa butter patches against DI water are shown. C. A typical interfacial deformation predicted by the theory, for 12 holes (in the same conditions as B), at $h_0=h_m-2.1r$. The defect circumference has been indicated as a black circle, the height of the undisturbed water line is indicated with a dashed green line and the meniscus height ($h_m$) is indicated with a dashed purple line.

The wetting experiments were also performed using a K12 tensiometer. The heterogeneous surfaces were immersed in a bath containing the aqueous EtOH solutions, as sketched in Figure 2A. The cumulative force generated by the deformation of the contact line to overcome the array of heterogeneities was measured at different submersion depths. The advancing contact line speed was controlled by raising the liquid container with the solution at 0.1 mm.s$^{-1}$. The liquid container diameter was sufficiently large to ensure that the meniscus at the container lateral wall did not affect the shape of the CL on the plate.

A typical force curve obtained experimentally is shown in Figure 2. A negative force corresponds to an upward force. Two peaks are visible in these curves, each associated with the wetting of one of the two rows of defects present on the surface. When the interface does not interact with any defect, the increasing buoyancy force on the submerged solid plate gives rise to an upward force linearly increasing with depth. This force, indicated with $f_{buoyancy}$ in Figure 2 should be subtracted to reveal the capillary force induced by the defects. The depinning force ($f_d$) and depth ($h_d$) were measured at the two peaks and the depinning energy ($E_d$) was calculated by integrating the force-depth curve. For comparison purposes the slopes of the curves ($\Delta f/\Delta h$) were computed by linearly fitting the force curves over the range $0.15 f_d - 0.85 f_d$.

The meniscus shape was measured by recording the submersion of the plates into excess liquid using a high-definition camera (Basler, Germany). Plates were affixed to a TA HD Texture Analyser (Stable Micro Systems, UK) and submerged at 0.1 mm.s$^{-1}$ to match the test protocol used on the K12 Tensiometer. The images obtained during the experiments performed with the Texture Analyser had a better quality compared to those obtained during the wetting experiments, because the contact line was at a constant height. Because the advancing liquid contact angle $\theta_{adv}$ was often >90°, the camera was tilted at an angle that facilitated the accurate recording of the meniscus on the s/s (typically an inclination of 45° was used). Where $\theta_{adv}$ was <<90° the incline was reduced to near 0°.

Figure 1 shows schematically the key variables describing the geometry in the *x/y* and *x/z* plane. The depression or rise of the contact line (distance in the z-direction) are measured with respect to the undisturbed meniscus (as y→∞). The meniscus rise on an undisturbed substrate, may be calculated from Equation 11, and is denoted by $h_m$:

$$h_m = \kappa^{-1}[2(1-\sin\theta)]^{1/2} \approx \cot\theta \cdot \kappa^{-1}$$

Equation 11

The approximated form of Equation 11 is valid for contact angles around 90°. The distance along the z coordinate between $h_m$ and the point of minimum CL depression is denoted $h_{max}$, as illustrated graphically in the third pane of Figure 1. In all instances, the contact angle of the liquid on the substrate is assumed to be $\theta \approx \theta_{adv}$. The last pane of Figure 1 shows the tangential depinning incline $\zeta$, in the x/z-plane at $x = r\sin\beta$ (discussed further in the analysis). $h_{max}$, $\beta$ and $\zeta$ were measured via image analysis from the images taken at three depths: $h_m - r$, $h_m - 2r$ and $h_d$ (the depinning depth), as schematically shown in Figure 8 D.

# Results and Discussion

## Depinning Forces

From the raw force data shown in Figure 2, one can observe that the pinning forces generated by holes in a stainless steel plate (blue peaks) are higher than the forces induced by the cocoa butter defects on the same plates (light brown peaks). Following the analysis described in the previous paragraph, the depinning force and energy for the different experiments described in Table 1 were calculated, and plotted in Figure 3 A and B. Figure 3 A shows that the depinning force scales linearly with the total pinning length calculated by multiplying the defect diameter and the number of defects. Figure 3 B shows that the depinning energy depends almost linearly on the total defect area.

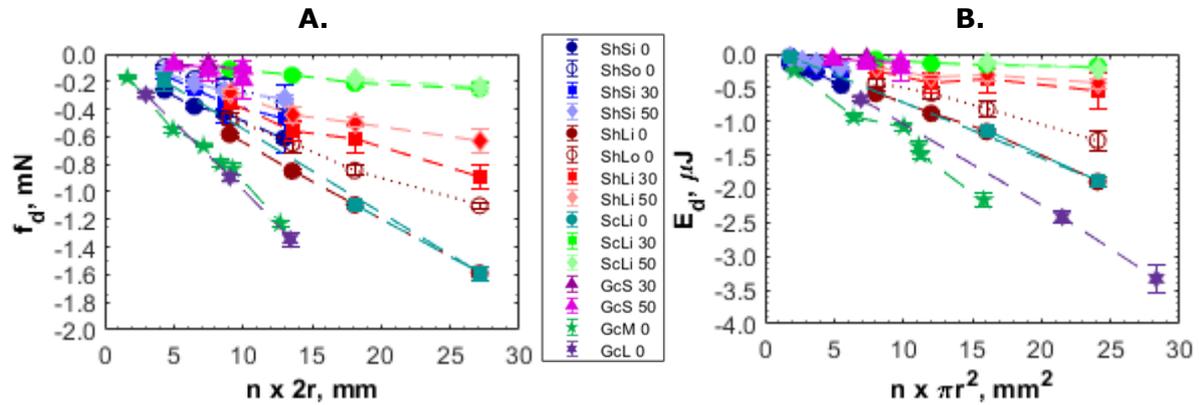

*Figure 3. A. Depinning force plotted against the product of the number of defects times their diameter for various defects (cocoa butter on stainless steel, holes in stainless steel, cocoa butter on glass) submerged into various aqueous EtOH solutions. B. Depinning energy plotted against the total defect surface. The labels are described in Table 1 and its caption. In the inset, the depinning force is plotted against the product of the number of defects times their diameter. Error bars indicate the standard deviation (5 repetitions).*

The linear dependence of the depinning force and energy can be interpreted by considering the change in spreading factor over the defect, already introduced in Eq.2. For an array of *n* defects, the depinning force and energy can be expressed as:

$$f_d = 2nr\gamma\Delta \cos\theta$$

Equation 12

$$E_d = n\pi r^2 \gamma\Delta \cos\theta$$

Equation 13

Figure 4 A and B shows a comparison of the experimental depinning force and energy, already shown in Figure 3, re-plotted against the expression proposed in Equation 12 and Equation 13, respectively. The lowest depinning forces and energies are measured with cocoa butter defects in the holes, while the highest with empty holes, consistently with the lower difference in spreading factors for fat defects.

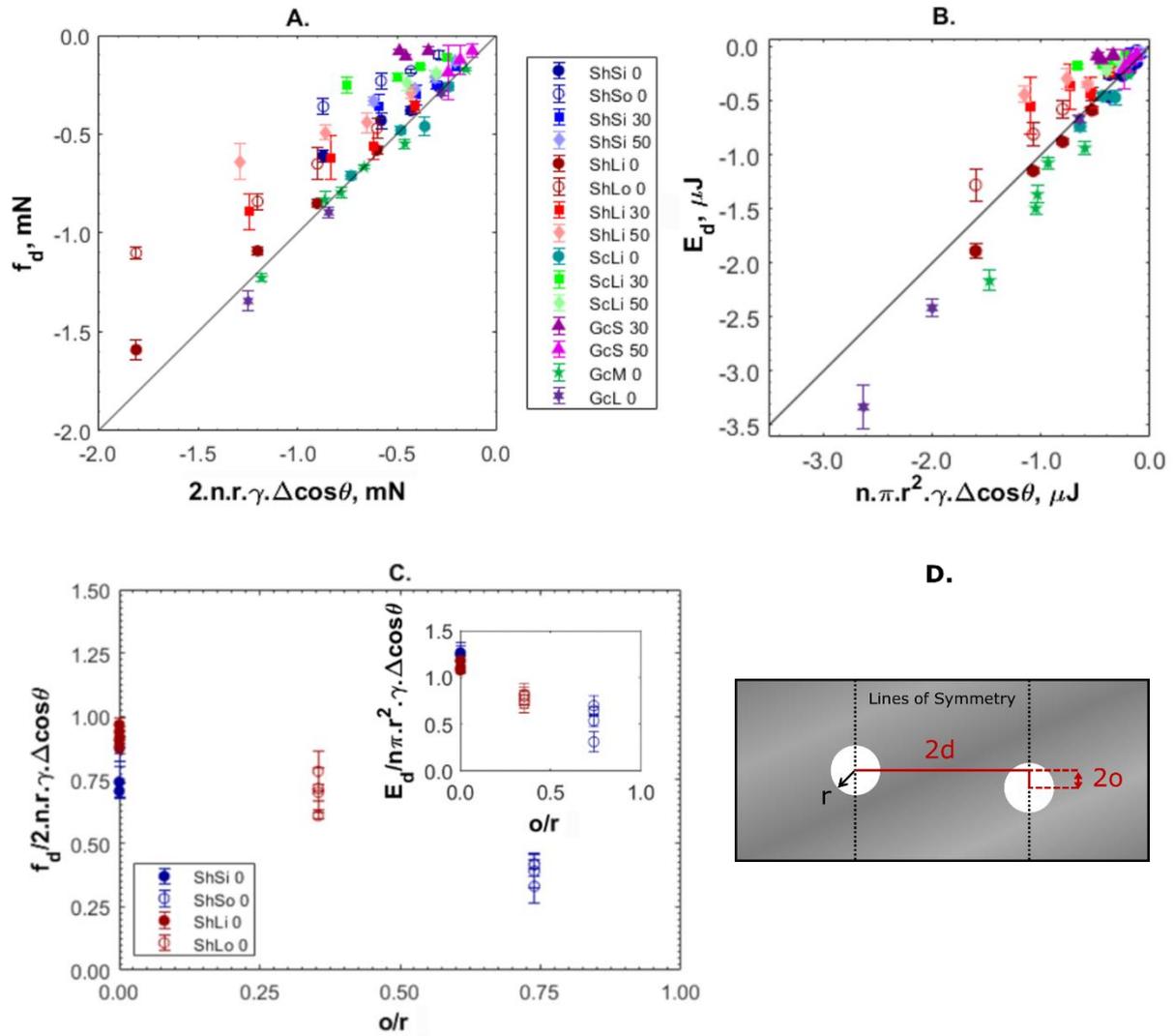

*Figure 4. A. Depinning force for various defects (cocoa butter on stainless steel, holes in stainless steel, cocoa butter on glass) submerged into various aqueous EtOH solutions. Hollow markers are used for surfaces with staggered (offset) defects. B. Depinning energy C. Dimensionless force vs the relative offset of the defects. The Inset in C shows the dimensionless energy vs the relative offset of the defects. The labels are described in Table 1 and its caption. D. Schematic diagram illustrating the offset of the holes in the x-z plane of the stainless steel plates.*

In general terms, the experimental data agree better to the scaling proposed in Equation 12 and Equation 13 when fewer defects are present, which corresponds to defects that are farther and acting independently from one another.

Furthermore, the agreement of the experimental data with the scaling proposed in Equation 12 and Equation 13 is better for the experiments performed using water, than for water-ethanol solutions, as shown in particular in terms of the depinning force.

Equation 12 was also not found to scale well with the data for the offset holes for both sizes (the open navy and burgundy data points in Figure 4), since the force and energy predicted by Equation 12 and Equation 13 are greater in magnitude than that observed experimentally. This is demonstrated in Figure 4 C and D. This discrepancy is greater for the smallest radius (0.27 mm), for which the experimental data were found to be 40% of the prediction, compared to 66% for the 0.56mm radius holes. The surfaces with staggered (offset) defects did not give rise to double force peaks, but rather to more flattened and elongated peaks. Indeed, the interface did not depin twice when wetting offset defects. The offset (o) considered in the study was 0.4 mm. The ratio between the offset and the defect radius was therefore o/r = 0.74 for r = 0.27 mm and o/r = 0.35 for r = 0.56 mm. For a high enough offset or small enough holes, the CL would encounter two rows of n/2 defects. Given that Equation 12 assumes that the *n* holes are aligned, the dimensionless force $f_d$ is expected to reach 0.5 for large o/r, as observed in figure 4C. Conversely, as o/r decreases, the normalised force is closer to unity, since the offset becomes negligible.

To test the effect of $\Delta cos\theta$ independently from the surface tension ($\gamma$), cocoa butter (cb) was used as a chemical heterogeneity, thus creating a 4-phase CL (steel, water, air, cocoa butter). Cocoa butter was found to decrease the depinning force and energy, (when compared to holes) for all EtOH concentrations. It was found that Equation 12 and Equation 13 were able to predict the experimental depinning force ($R^2$ = 0.95) and energy ($R^2$ = 0.84) for the cocoa butter defects on stainless steel submerged into water.

A stronger over prediction of depinning force and energy was observed as the liquid EtOH concentration increased, resulting in a worse agreement with Equation 12 and Equation 13. Two explanations can be proposed for this over prediction: i) Ethanol is a solvent for molten cocoa butter and it is speculated that a partial solubilisation of the liquid fraction of cocoa butter in EtOH solutions could change the interfacial properties; ii) The simple scaling of Equation 12 and Equation 13 does not take into account surface roughness of the cocoa butter and the effects this may have on wetting. Unlike other chemical heterogeneities, the surface of cocoa butter can become rough due to fat crystallisation[11, 38-41] and this phenomenon is not accounted for in Equation 12 and Equation 13. Many authors have discussed and investigated the increase in hydrophobicity as a result of surface roughness[31, 42-44]. Surface roughness can increase hydrophobicity for cocoa butter surfaces wetted by water ($\theta$ ~120°) and hydrophilicity (solvophilicity) for cocoa butter surfaces wetted by 50% EtOH solutions ($\theta$ ~70°).

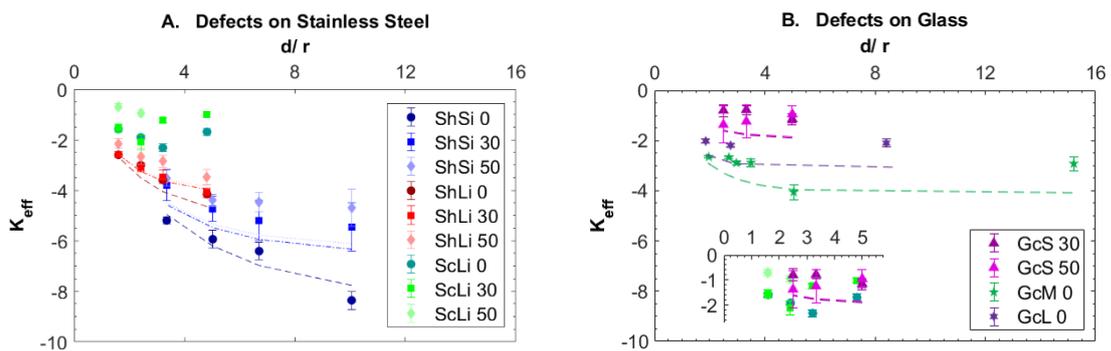

*Figure 5. The effective dimensionless interfacial stiffness, $K_{eff}$, for defects on: A. stainless steel and B. glass. $K_{eff}$ is defined in Equation 14. The inset in B shows only the cocoa butter samples of similar size on stainless steel and glass. The labels are described in Table 1. Error bars indicate standard deviation (5 repetitions), dotted lines indicate the stiffness predicted by the theory.*

As already discussed, the force induced by the interface deformation in presence of defects, $\Delta f$, scales approximately linearly with the penetration depth of the defects into the liquid, $\Delta h$. A linear fit in the range 15-85% of the peak force was

applied, in order to introduce an effective dimensionless interfacial stiffness (per defect) as:

$$K_{eff} = \frac{\Delta f/(n\,r\,\gamma)}{\Delta h\,\kappa}$$

Equation 14

Figure 5 shows the values of $K_{eff}$ for the different experimental conditions considered. Holes of similar $d/r$, but different $r$ lead to different results, indicating a dependence of $\Delta F/\Delta H$ on $r$. The theory outlined above compares favourably with the experimental data, save for an occasional small over prediction. The theory is less predictive at higher d/r and smaller r though this may be due to stronger boundary effects or to the limited accuracy of force measurements. In general, $K_{eff}$ increases as the periodicity decreases indicating neighbouring defects perturb less the interface and reduce the force required for contact line depinning. The inset in Figure 5 B shows $K_{eff}$ for cocoa butter defects of comparable size on stainless steel (r = 0.56 mm) and on glass (r = 0.62 mm). The curves overlap indicating that the stiffness of the interface is defect-induced and independent of the substrate.

Although the theory assumes pinning at the base of the defect, its predictions agree well also with the results obtained with cocoa butter on glass (green pentagrams and purple hexagrams in Figure 5). It appears that as ΔS increases, the ability of the theory to predict the pinning force increases. This may explain the good agreement between the theory and the cocoa butter on glass experiments, in contrast to the much worse agreement for cocoa butter on steel (where $\Delta\theta$ is ~12°). In the latter case, the 4-phase CL greatly decreased the force, which may be caused by the ability of the CL to de-pin in sections and move on top of the defects, thus reducing the meniscus depression and the resulting force.

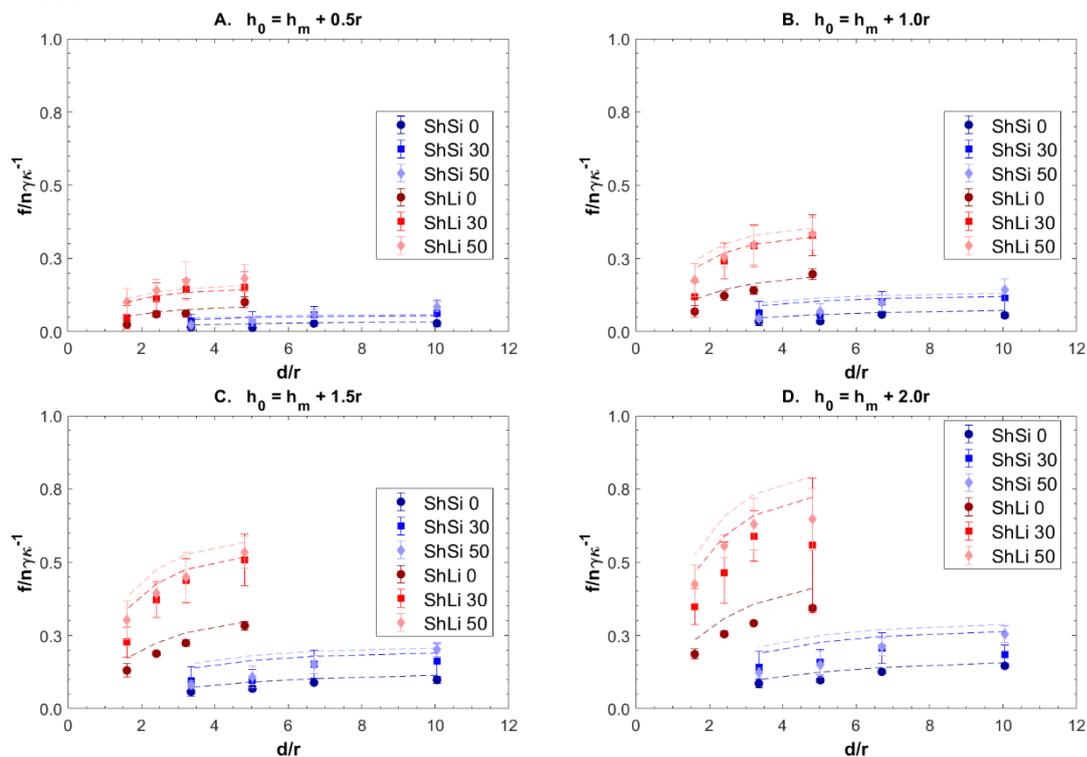

Figure 6. Dimensionless force induced by the deformation of the interface obtained for holes on stainless steel vs the dimensionless hole periodicity. The holes were submerged at four depths into

*0, 30 & 50% EtOH solutions. Dashed curves show the predictions of the theory. Error bars indicate standard deviation determined from 5 repetitions. The labels are described in Table 1.*

Figure 6 shows a comparison of the force induced by the defects measured in experiments and the force predicted by the theory. Subfigures A, B, C and D consider respectively immersion depths beneath the meniscus height of 0.5, 1, 1.5 and 2 times the defect radius. Both hole sizes and different ethanol/water solutions are considered. The theory offers the ability to predict pinning forces for different defect spacing (d) and various depths. The theory captures well also the interaction of defects in terms of interface deformation and depinning force, which cannot be captured by the scaling proposed in Equation 12 and Equation 13. The theory, however, does not allow a prediction of the depinning position itself.

## Interface shape

An example of a theoretical prediction of the contact line shape and water air interface shape as deformed by a hole of r = 0.56 mm at defect distances of 1.8 mm at a depth of $h_m$ - 2.1r was provided in Figure 2 C.

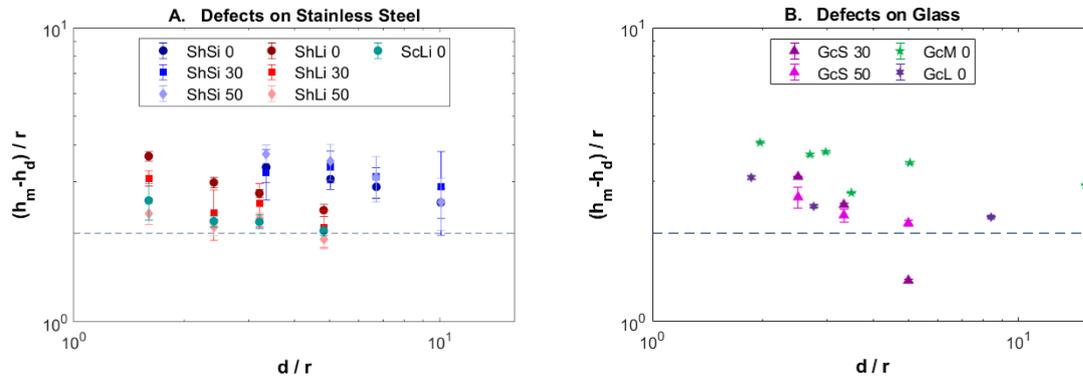

*Figure 7. Difference between meniscus height ($h_m$) and the depinning height ($h_d$) for defects on stainless steel (A) and on glass (B) submerged into various EtOH concentrations vs the ratio of periodicity divided by radius. The labels are described in Table 1. Dashed blue line indicates when $h_m$ - $h_d$ = 2r. Error bars indicate standard deviation (n=5).*

Figure 7 illustrates the difference between the depinning depth and the undisturbed meniscus height for various defects and liquid compositions. The defect radius (*r*) was varied in the range 0.26-1.44mm and *d/r* in the range 1.6-15.2. As *d/r* increases, $h_m - h_d$ decreases for all curves, i.e. the depinning depth decreases. The effect of d/r is more pronounced for bigger defects than for the smaller defects. Decreasing $\Delta S$ between the defect and the substrate was not seen to impact significantly $h_d$ for the smallest defects, probably due to the limitations of the force measurement, while as $\Delta S$ decreased, $h_m - h_d$ decreased for the biggest defects on steel and the samples comprising cocoa butter defects on glass. At $d/r = 1$, the defects would touch at their edges, creating a periodic defect line of circular segments. As $d/r \to \infty$, and $2r < h_{m,defect}$, it appears as though $(h_m - h_d) \to 2r$, as indicated in Figure 7 with a dashed line. For the cocoa butter data on s/s (r = 0.56mm), the ($h_m$ - $h_d$) are close to 2r. This can be partially explained by the change in meniscus height being roughly equal to the radius of the defect and that the CL is able to move onto the surface of the defect, without staying pinned at the edge. Where $r > h_{m,defect} - h_{m,substrate}$, $h_m - h_d$ could be lower than $2r$ as the contact line can de-pin part way over the defect. For cocoa butter on glass, the behaviour is similar to holes in stainless steel where, as $d/r \to \infty$,

($h_m - h_d$) → 2r. This can be explained in terms of similarly large ΔS. It was not possible to accurately measure the depinning height of the cocoa butter defects submerged into EtOH solutions since no clear depinning position was visible on the images.

Figure 8 A, B and C show the minimum distance of the contact line depression from the position of the undisturbed meniscus on the substrate ($h_m$), i.e. the highest position of the depressed meniscus ($h_{max}$), as defined in Figure 1. The depth of the defects ($h_0$) is measured from the undisturbed water interface height, to the base of the defect as illustrated in Figure 1. The contact line depression was investigated at three depths, $h_0 = h_m - r$, $h_0 = h_m - 2r$ and at depinning, $h_0 = h_d$ as shown in Figure 8 A, B and C respectively. The minimum depression of the CL can only occur between the base of the hole ($h_0$) and the meniscus height of the substrate ($h_m$). As such, $h_{max}$=0 indicates that the CL returns to the meniscus height of the substrate without defects. Conversely, $h_{max}/(h_o-h_m)$=1 indicates a straight contact line at the base of the defect.

Assuming the contact line remains pinned in the direction of submersion at the defect edge, the remains pinned across the plate leading to $h_m + r < h_{max} < h_d$. As → ∞ , $h_{max} → 0$ as the impact of the defect ceases to influence the CL. The capillary length has been indicated in the figures as a visual comparison to the length of deformation solely by the defect. Other authors[26] discussed that as d becomes greater than the capillary length, synergistic effects on the contact line should not be observed however they also discuss similar cases to those presented where this does not hold true.

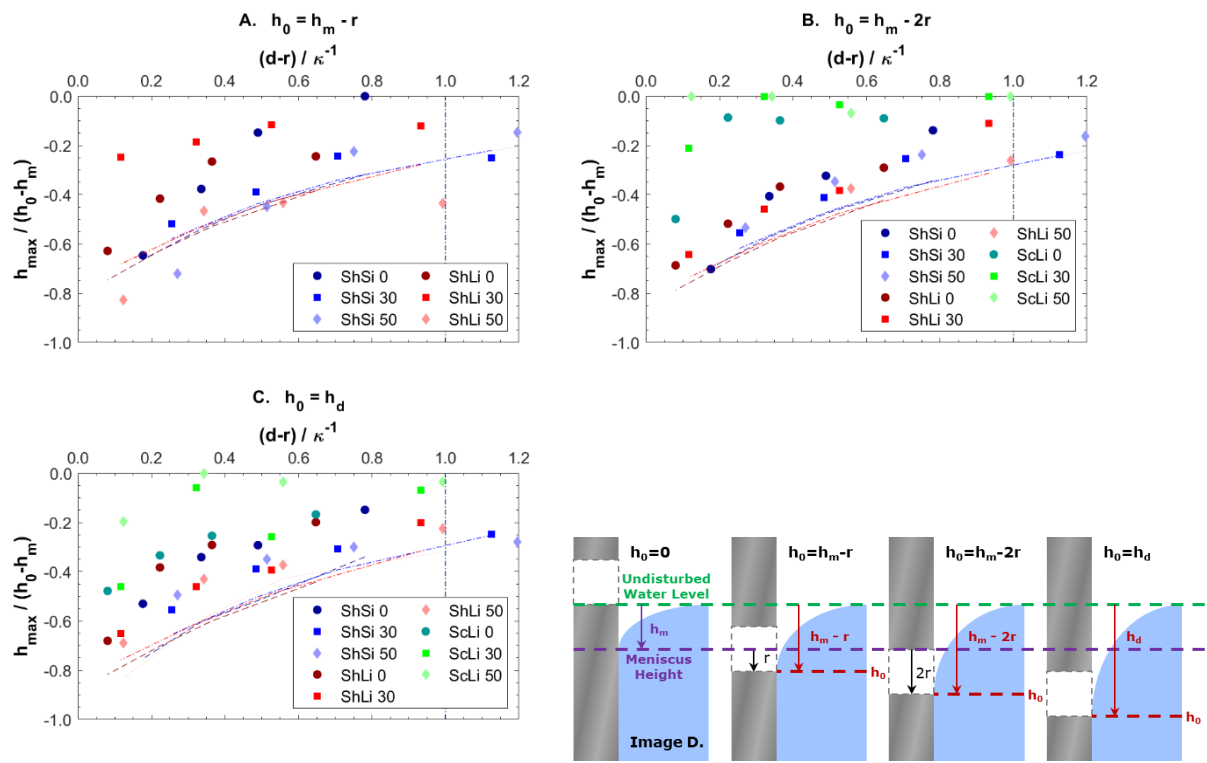

Figure 8. Minimum depression of the meniscus ($h_{max}$) on stainless steel (ss) at various $h_0$ (A. $h_0 = h_m - r$, B. $h_0 = h_m - 2r$, C. $h_0 = h_d$) shown over a range of periodicities between defects (d-r). Model data for $h_d$ was obtained by considering $h_d$ obtained from experimental data. The labels are described in Table 1. Dash dotted blue line indicates a distance of the capillary length ($\kappa^{-1}$) from the

*edge of the defect. The schematic diagram in D is provided to visualise the position of the defect relative to the undisturbed, horizontal water surface.*

Experimental results for holes in stainless steel appear to collapse to a master curve for both sizes irrespective of ethanol concentration. This is more apparent at the depinning depth and at a depth $h_m$-2r, rather than at $h_m$-r, which has been attributed to the variability in the test method. As the periodicity decreased, the depression of the contact line increased relative the base of the hole ($h_{max}/(h_0-h_m)$). This would suggest that the meniscus depression is largely dependent on the depth and distance between the defects. The introduction of cocoa butter greatly decreased the depression of the meniscus, consistently to the ability of the CL to move on the defects.

Theoretical predictions for the CL shape at the depinning position were calculated by using the averaged depinning depths measured experimentally as inputs. As the collapse of the theoretical lines in Figure 8 shows, $h_{max}$ is proportional to $h_0 - h_m$ independent of the defect depth ($h_0$), defect radius as well as $\Delta S$. The effect of $\Delta S$ is not strong, in the conditions considered.

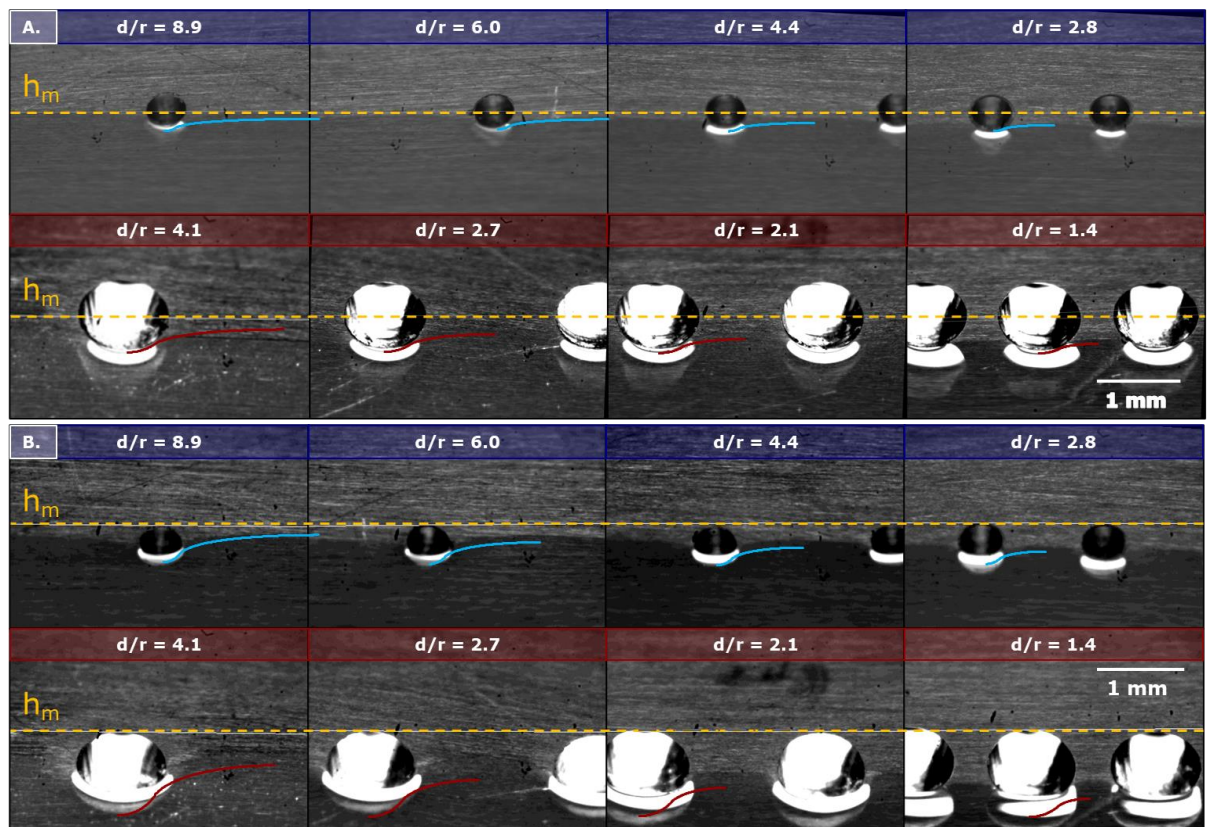

*Figure 9. Images of a plate with holes submerged into water at the points A. $h_0 = h_m - r$ and B. $h_0 = h_m - 2r$. The theoretical prediction of the meniscus shape has been superimposed for direct comparison. The height of the meniscus ($h_m$) has been indicated on the images with a dashed orange line. The red and blue colours reflect the colours used for the small and large defects in the previous figures.*

Figure 9 shows pictures of the meniscus shape at $h_0 = h_m - r$ and $h_0 = h_m - 2r$. The meniscus shape obtained from the theory has been superimposed for direct comparison. In all instances the theory overpredicts the CL depression, although the agreement is generally good. The theory is closer to the experiments at smaller inclines, which reflects the assumption of small meniscus slopes made in the theory.

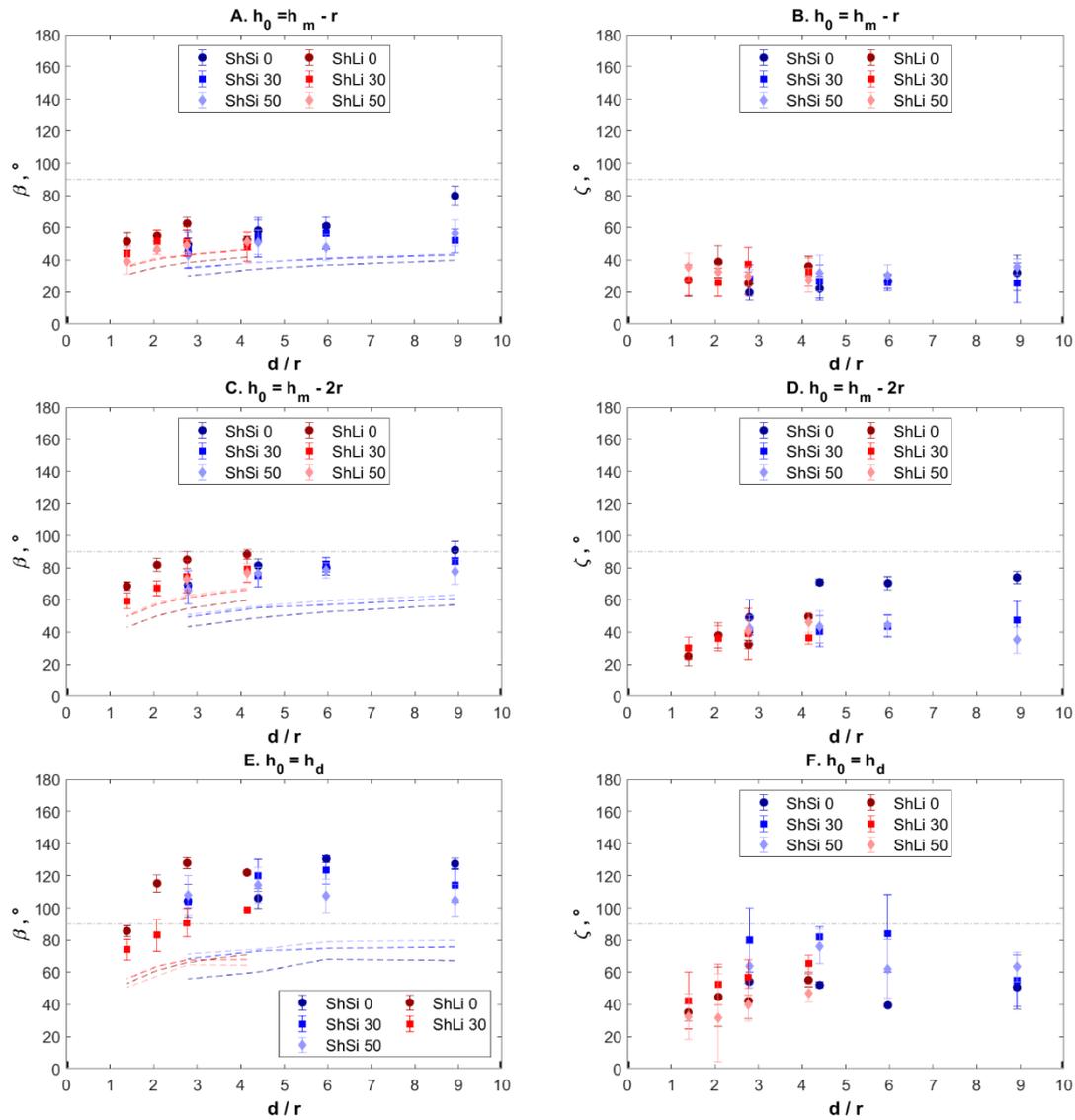

Figure 10. Depinning angle, β, on the x,z plane (A,C,E, on the left) and the tangential depinning angle, ζ, of CL from the defect on the x,z plane (B,D,F, on the right) plotted against the ratio periodicity/radius. A&B $h_0 = h_m - r$, C&D, $h_0 = h_m - 2r$, E&F $h_0 = h_d$. The labels are described in Table 1. Dashed lines show model predictions. Dotted-dashed show 90° angles. Error bars indicate standard deviation (n=3).

The shape of the deformed interface can be analysed in more details, by considering the β and ζ angles in the x/z plane, as defined in Figure 1. Figure 10 A., C., E., show the experimental and theoretical depinning angle, β, at $h_0 = h_m - r$, $h_0 = h_m - 2r$ and $h_0 = h_d$ respectively. As $h_0$ increases, so does β. Increasing d/r increases β slightly, but the radius was not found to affect β directly as it can be seen by the similar curves obtained with the two defect sizes. Changes in ΔS were also not seen to have a large influence on β. In all instances the theory under-predicts the depinning angle. Indeed, the current formulation of the model assumes that β<90°, while experimental results show that depinning in several cases occurs at β>90°, thus justifying the gap observed between theory and experiments, at larger depths.

Figure 10 B., D., F., show the apparent tangential depinning angle, ζ. This angle is observed experimentally to be always lower than β. The apparent angle ζ cannot be explained by the theory and may indicate some form of incomplete pinning on the circumference of the hole, suggesting possibly a partial filling of the holes in the y direction. Such partial filling, however, was not visible in the videos taken during the experiments. As a result, these apparent ζ angles are not fully understood, as elaborated in the conclusions. Unsurprisingly, at higher depth, ζ increases, which would be compatible with an increasingly incomplete pinning. A similar behaviour is observed for both hole sizes.

## Conclusions

In this work, the deformation of an advancing contact line and the force caused by an array of surface heterogeneities have been studied experimentally. The results were compared with a simple theory valid under the assumption of small interfacial inclinations and perfect pinning at the bottom of the defects. The theory correctly reproduces the trends observed in experimental data at small immersion depths, while a simple scaling has also been shown to reasonably predict the depinning force and energy. Unlike the simple scaling, the model is able to reproduce the trends observed in the experimental data, up to the point of depinning, not just at the depinning position. The theory also accounts for the interaction of the deformations induced by nearby defects, which cannot be captured by the simple scaling considering isolated defects. Interacting defects reduce the overall pinning force, this interaction decreases as d/r increases up to a point where defects act independently as described by the simple scaling.

Defects such as cocoa butter patches on stainless steel, causing a 4-phase CL, resulted in a lower dimensionless depinning force, when compared to those leading to a 3-phase CL. On stainless steel, the 4-phase contact lines were seen to move on top of the defects, leading to the reduction in pinning forces, which lead to smaller interfacial deformations. Where the difference in spreading factors of substrate and defects, $\Delta S$, is higher, the depinning force and interfacial deformation were more akin to those of a 3-phase CL and the theory matched better the experiments. The depinning energy was generally found to scale well with $\pi r^2 n. \Delta S$.

The depinning angle, $\beta$, predicted by the theory compared favourably with experiments at small submersion depths. However, a tangential depinning angle, ζ, was also observed experimentally, suggesting an incomplete pinning of the interface at the bottom of the defect. The depinning angles increased strongly with depth, but did not depend strongly on the liquid contact angles, on the defect radius and distance. Further work is needed to better understand the interface shape and to confirm the origin of the observed tangential depinning angle, ζ.

These results can help in understanding better the effect of surface defects on spontaneous and forced wetting, thus enabling to better engineer wetting properties.

# Appendix: Numerical method

Our numerical method represents the interface shape as the series

$$H(X,Y) = \sum_{i=0}^{\infty} A_i \cos\left(\frac{\pi i}{d} X\right) \exp\left[-(1 + \pi^2 i^2/d^2)^{1/2} Y\right]$$

where the coefficients $A_i$ are to be determined to satisfy the boundary conditions on the wall, $Y = 0$. (Note that the solution above has been chosen as the most general combination of separable solutions of the linear Laplace-Young equation respecting the underlying symmetry and behaviour as $Y \rightarrow \infty$.)

The coefficients $A_i$ are determined by truncating the series after $N + 1$ terms and discretizing the interval [0,d) with $N + 1$ equally-spaced points. For a given value of the depinning point $\beta$, the relevant boundary conditions can be evaluated at each point, thereby giving an $(N + 1) \times (N + 1)$ system that can readily be solved numerically (e.g. in MATLAB). For a given hole position, the value of $\beta$ is determined by requiring that the effective contact angle at the inner edge of the hole is that on the plate itself.

# Conflicts of Interest

There are no conflicts of interest to declare.

# Acknowledgments

The Authors would like to thank the Engineering and Physical Sciences Research Council (UK) for funding. SM would also like to thank the Erasmus+ program for partially funding a secondment at INRAe/AgroParisTech Massy.